\documentclass{iaus}
\usepackage{graphicx}

\title[Predicting the Yields of Photometric Surveys]
{Predicting the Yields of Photometric Surveys for Transiting Planets}

\author[Thomas G. Beatty]
{Thomas G. Beatty$^1$}

\affiliation{$^1$ Department of Physics and MIT Kavli Institute, Massachusetts Institute of Technology, 77 Massachusetts Ave., Cambridge, MA 02139 \\ email: {\tt tbeatty@space.mit.edu}}

\pubyear{2008}
\volume{253}
\jname{Transiting Planets}
\editors{D. Sasselov \& D. Queloz}

\begin{document}

\maketitle

\begin{abstract}
Observing extrasolar planetary transits is one of the only ways that we may infer the masses and radii of planets outside the Solar System. As such, the detections made by photometric transit surveys are one of the only foreseeable ways that the areas of planetary interiors, system dynamics, migration, and formation will acquire more data. Predicting the yields of these surveys therefore serves as a useful statistical tool. Predictions allows us to check the efficiency of transit surveys (``are we detecting all that we should?'') and to test our understanding of the relevant astrophysics (``what parameters affect predictions?''). Furthermore, just the raw numbers of how many planets will be detected by a survey can be interesting in its own right. Here, we look at two different approaches to modeling predictions (forward and backward), and examine three different transit surveys (TrES, XO, and Kepler). In all cases, making predictions provides valuable insight into both extrasolar planets and the surveys themselves, but this must be tempered by an appreciation of the uncertainties in the statistical cut-offs used by the transit surveys. 
\keywords{methods: numerical, planets and satellites: general, surveys, techniques: photometric}
\end{abstract}

\firstsection

\section{Introduction}

Planetary transits are one of the many methods by which extrasolar planets
have been discovered, and are also the one that provides the most complete set
of information about the planetary system. Only planets with very
specific orbital characteristics have a transit visible from Earth,
because the orbital plane has to be aligned to within a few degrees of
the line of sight. Therefore transiting planets are
rare. Nevertheless, a transiting extrasolar planet offers the
opportunity to determine the mass of the planet - when combined with radial velocity (RV)
measurements - since the inclination is now measurable, as well as the
planetary radius, the density, the composition of the planetary
atmosphere, the thermal emission from the planet, and many other
properties (see \cite[Charbonneau et al. (2007)]{charbonneau2007} for a review). Additionally,
and unlike RV surveys, transiting planets should be readily detectable
down to $1 R_\oplus$ and beyond, even for relatively long periods.

Having accurate predictions of the number of detectable transiting
planets is immediately important for the evaluation and design of
current and future transit surveys. For the current surveys,
predictions allow the operators to judge how efficient are their
data-reduction and transit detection algorithms. Future surveys can
use the general prediction method that we describe here to optimize
their observing set-ups and strategies. More generally, such
predictions allow us to test different statistical models of
extrasolar planet distributions. As more transiting planets are discovered, these statistical properties are increasingly becoming the frontier of research, shifting the focus of the field away from individual detections.

\section{The Failure of Simple Estimates}

Using straightforward estimates it appears that observing a planetary
transit should not be too difficult, presuming that one observes a
sufficient number of stars with the requisite precision during a given
photometric survey. Specifically, if we assume that the probability of
a short-period giant planet (as an example) transiting the disk of its
parent star is 10\%, and take the results of RV surveys which indicate
the frequency of such planets is about 1\% (\cite[Cumming et al. 2008]{cumming08}),
together with the assumption that typical transit depths are also
about 1\%, the number of detections should be $\approx 10^{-3}N_{\leq
1\%}$, where $N_{\leq 1\%}$ is the number of surveyed stars with a
photometric precision better than 1\%.

Unfortunately, this simple and appealing calculation fails. Using this
estimate, we would expect that the TrES survey, which has examined
approximately 70,000 stars with better than 1\% precision, to have
discovered 70 transiting short period planets. But, at the date of
this writing, they have found four. Indeed, overall only 51 transiting
planets have been found at this time by photometric surveys
specifically designed to find planets around bright stars\footnote{As
of June 2008. See the Extrasolar Planets Encyclopedia at
http://exoplanet.eu for an up-to-date list.}. This is almost one
hundred times less than what was originally predicted by somewhat more
sophisticated estimates (\cite[Horne 2003]{horne2003}).

Clearly then, there is something amiss with this method of estimating
transiting planet detections. Several other authors have developed
more complex models to predict the expected yields of transit
surveys. \cite[Pepper at al. (2003)]{pepper2003} examined the potential of all-sky surveys,
which was expanded upon and generalized for photometric searches in
clusters \cite[Pepper \& Gaudi (2005)]{pepper2005}. \cite[Gould et al. (2006)]{gould2006} and \cite[Fressin et al. (2007)]{fressin2007} tested whether the OGLE planet detections are statistically consistent
with radial velocity planet distributions. \cite[Brown (2003)]{brown2003} was the
first to make published estimates of the rate of false positives in
transit surveys, and \cite[Gillon et al. (2005)]{gillon2005} model transit detections to
estimate and compare the potential of several ground- and space-based
surveys.

As has been recognized by these and other authors, there are four
primary reasons why the simple way outlined above of estimating
surveys yields fails.

First, the frequency of planets in close orbits about their parent
stars (the planets most likely to show transits) is likely lower than
RV surveys would indicate. Recent examinations of the results from the
OGLE-III field by \cite{gould2006} and \cite{fressin2007} indicate
that the frequency of short-period Jovian-worlds is on the order of
$0.45\%$, not $1.2\%$ as is often assumed by extrapolating from RV
surveys \cite{marcy2005}. \cite{gould2006} point out that most
spectroscopic planet searches are usually magnitude limited, which
biases the surveys toward more metal-rich stars, which are brighter
at fixed color. These high metallicity stars are expected
to have more planets than solar-metallicity stars \cite{santos2004,
fischer2005}.

Second, a substantial fraction of the stars within a survey field that
show better than 1\% photometric precision are either giants or early
main-sequence stars that are too large to enable detectable transit
dips from a Jupiter-sized planet \cite{gould2003, brown2003}.

Third, robust transit detections usually require more than one transit
in the data. This fact, coupled with the small fraction of the orbit a
planet actually spends in transit, and the typical observing losses at
single-site locations due to factors such as weather, create low
window probabilities for the typical transit survey in the majority of
orbital period ranges of interest \cite{vonbraun2007}.

Lastly, requiring better than 1\% photometric precision in the data is
not a sufficient condition for the successful detection of transits:
identifiable transits need to surpass some kind of a detection
threshold, such as a signal-to-noise ratio (S/N) threshold. The S/N of
the transit signal depends on several factors in addition to the
photometric precision of the data, such as the depth of the transit
and the number of data points taken during the transit
event. Additionally, ground-based photometry typically exhibits
substantial auto-correlation in the time series data points, on the
timescales of the transits themselves. This additional red noise,
which can come from a number of environmental and instrumental
sources, substantially reduces the statistical power of the data
\cite{pont2006}.

\section{Approaches to Modeling}

There are, generally speaking, two different approaches to the problem of statistically modeling transit surveys: forward and backward modeling. Forward modeling is the more general of the two, since it strives to predict the number of detections a survey should see by working ``forward'' from the parameters of the survey telescopes and target fields. As a result, forward modeling schemes are dependent not only on the uncertainties in extrasolar planet statistics, but also on the uncertainties in the stellar and galactic properties that must be used to model the observed star field. 

Backwards modeling avoids these uncertainties, at the loss of generality, by taking a known set of stars and survey parameters. This differs from forward modeling in that the properties of the target stars are known \emph{a priori}, usually in the form of an input catalog or something similar. This knowledge considerably increases the accuracy of the predictions compared to the forward modeling case, since one no longer needs to be concerned with statistically modelling the target stars. Nevertheless, only a few transit surveys, planned or operating, have devoted the telescope time towards constructing thorough input catalogs. 

\subsection{Forward Modeling}

Thinking about the forward modeling problem in the most general sense, we can describe the average number of planets that a transit survey should detect as the probability of
detecting a transit multiplied by the local stellar mass function,
integrated over mass, distance, and the size of the observed field
(described by the Galactic coordinates $(l,b)$:

\begin{equation}\label{eq:10}
\frac{d^6 N_{det}}{dR_p\ dp\ dM\ dr\ dl\ db} = \rho_*(r,l,b)\ r^2 \cos b\ \frac{dn}{dM}\ \frac{df(R_p,p)}{dR_p\ dp}\ P_{det}(M,r,R_p,p),  
\end{equation}
where $P_{det}(M,r,R_p,p)$ is the probability that a given star of
mass $M$ and distance $r$ orbited by a planet with radius $R_p$ and
period $p$ will present a detectable transit to the observing
set-up. $\frac{df(R_p,p)}{dR_p\ dp}$ is the probability that a star
will possess a planet of radius $R_p$ and period $p$. $dn/dM$ is the
present day mass function in the local solar neighborhood, and
$\rho_*$ is the local stellar density for the three-dimensional
position defined by ($r,l,b$). We use $r^2 \cos b$ instead of the
usual volume element for spherical coordinates, $r^2 \sin \phi$,
because $b$ is defined opposite to $\phi$: $b=90^\circ$ occurs at the
pole.

As an example of forward modeling, \cite{beatty2008} (hereafter B\&G) simulated the TrES and XO transit surveys under various magnitude and signal-to-noise (S/N) limits using the planet frequencies of \cite{gould2006}. We found that both surveys have discovered about the number of transits that one would expect, though the exact numbers are highly sensitive to the assumed magnitude and S/N cut-offs, as well as to the amount of red-noise. 

For the TrES survey, we simulated 13 of the TrES target fields. The locations and observation times for each was collected from the Sleuth observing website\footnote{http://www.astro.caltech.edu/$\sim$ftod/tres/sleuthObs.html}. Figure 1 shows the expected distribution of detections, in terms of host star mass and radius, for the Lyr1 field under three different scenarios. Lyr1 is the field containing TrES-1, and is adjacent to the field containing TrES-2; both stars are marked in the figure, and both lie close to the predicted maximum number of detections. 

\begin{figure}[h]
\begin{center}
\includegraphics[width=3.5in]{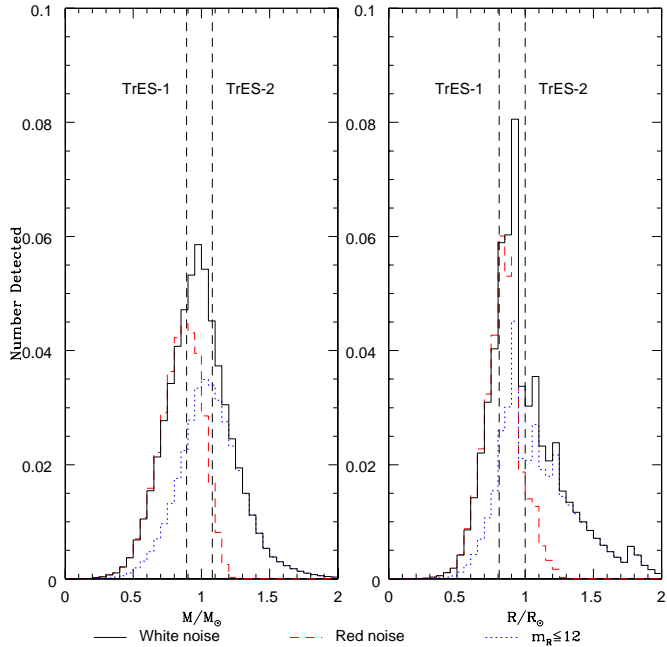}
\caption{Predicted TrES detections in the Lyr1 field. The three scenarios demonstrate the effect that uncertainties in the magnitude cut-off and amount of red-noise can have on survey predictions.}
\label{fig1}
\end{center}
\end{figure}

In terms of raw numbers, B\&G find that with cut-offs of $m_R\leq13$ and S/N $\geq30$, TrES should detect 13 planets, as compared to the 4 planets that have been (as of June 2008) found. We selected these cut-offs as those used by the TrES team in their publications. Interestingly, however, none of the TrES planets have host stars dimmer than $m_R=12$. When we re-simulate the TrES fields using a limit of $m_R\leq12$, the number of detections dropped to 8.16. Alternatively, if we keep the $m_R\leq13$ limit - but add in 3 mmag of red-noise (\cite[Pont et al. 2006]{pont2006}) - the predicted number of detections drops again, to 7.68.

All of which serves to demonstrate the sensitivity of survey predictions to the statistical cut-offs used. 

In the case of XO, which uses a drift-scanning technique, B\&G simulated each of the six XO fields, spaced evenly in right ascension around the sky. The first planet discovered by the XO survey, XO-1b, (\cite[McCullough et al. (2006)]{mccullough2006}) is located in the field at 16 hrs, and so Figure 2 plots the distribution of detections within this field over the masses and radii of the host stars.

\begin{figure}[h]
\begin{center}
\includegraphics[width=3.5in]{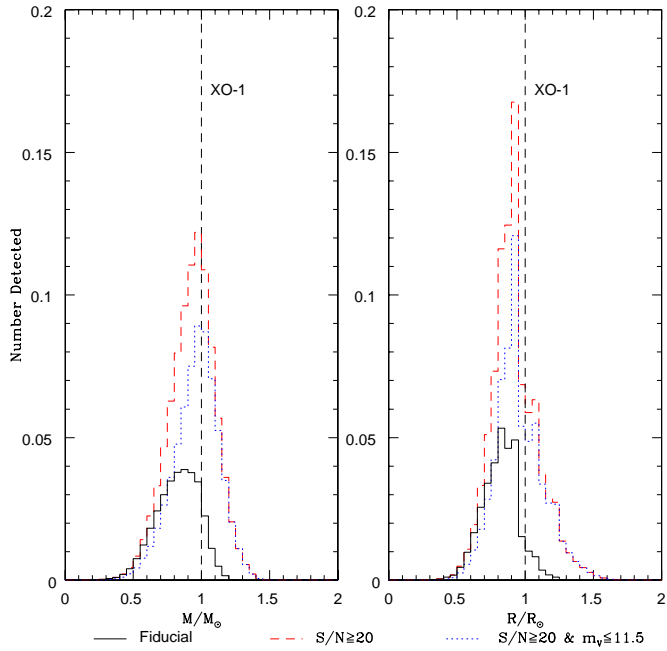}
\caption{Predicted XO detections in the survey's field at RA = 16 hrs. As with the TrES survey example, several different cut-off scenarios are shown.}
\label{fig2}
\end{center}
\end{figure}

Similar to TrES, Figure 2 shows several different cut-off scenarios. The fiducial case examined in B\&G is for  $m_V\leq12$ and S/N $\geq30$, and yields 2.02 predicted detections over all fields (of which only one is displayed in Figure 2). XO has actually detected 3 planets. If the S/N limit is lowered to S/N $\geq20$, the number of predicted detections increases to 5.86. Also similar to TrES, the actual magnitude limit of the XO survey may be brighter than what is in the literature. Indeed, all three XO host stars are at $m_V\leq11.5$ (XO-1, the dimmest, is at $m_V=11.3$). In the case of $m_V\leq11.5$ and S/N $\geq20$, the predicted detections are 4.21.

Again, all of which is to show that not only is it possible to make reasonable predictions for the number of planets that a photometric survey will find using a forward modeling approach, but also to illustrate how much those predictions, and by extension the real numbers, depend upon the statistical cut-offs used by the transit surveys.

\subsection{Backwards Modeling}

In the case that a survey has a known set of target stars, it is possible to use the properties of those stars directly in the simulations of the survey, and work ``backwards'' from that starting point to a set of predicted detections. Therefore, and unlike Equation 3.1, for backwards modeling the statistical treatment of the target stars is abandoned, and replaced by a direct summation over all the target stars,

\begin{equation}\label{eq:20}
\frac{d^2 N_{det}}{dR_p\ dp} = \sum_i \frac{df(R_p,p)}{dR_p\ dp}\ P_{det}(M_i,r_i,R_p,p).  
\end{equation}
Here we use the same definitions as Equation 3.1, but sum over $i$ targets.

Kepler is one of the very few surveys that has put together a comprehensive input catalog of all the stars in their target field. I was able to use the Kepler Input Catalog (KIC) as a demonstration of backwards modeling after David Latham kindly provided access to an early version of the KIC for use in these predictions. The target stars were pulled from the KIC by selecting all the entries with a spectroscopic $T_{eff}\leq10,000$K, $m_{Kepler}\leq15$, and spectroscopic $log(g)\geq3.5$. These criteria yielded 109,400 targets. I then used the mass-temperature and mass-radius relations from B\&G to estimate masses and radii for all of the selected stars. Table 1 shows the spectral distribution of the targets from the KIC, as well as the distribution predicted by B\&G using forward modeling. 

\begin{table}[h]
\begin{center}
\caption{Spectral Distribution of the Kepler Field}
\label{tab1}
{\scriptsize
\begin{tabular}{cccccc|c}
\hline 
& A & F & G & K & M & All\\  
\hline
   KIC: & 3549 &  55720&  45216&   4735&   180&  109400\\
B\&G:   & 5066 &  41094&  45568&  13351&   942&  106342\\
\hline
\end{tabular}}
\end{center}
\end{table}

Compared to ground-based transit surveys, Kepler will be able to discover many more close in giant planets. B\&G divides these planets into Very Hot Jupiters (VHJs) and Hot Jupiters (HJs) depending on their orbital period (1-3 and 3-5 days, respectively). Using the planet frequencies from \cite{gould2006},  Table 2 shows the results of using the KIC to predict the number of VHJs and HJs that Kepler will discover, and also the results of the similar forward modeling simulations done in B\&G. The numbers are very similar, which reflects the similarity of the total star counts in both the KIC and B\&G: Jupiter-sized planets are so easy for Kepler to detect, it gets all of them. The only question is exactly how many target stars Kepler will observe.

\begin{table}[h]
\begin{center}
\caption{Kepler's Transiting VHJs (1-3 days) and HJs (3-5 days)}
\label{tab1}
{\scriptsize
\begin{tabular}{cccccc|c}
\hline
 & A & F & G & K & M & All\\  
\hline 
VHJs:& 1.31& 16.35& 10.31& 0.88& 0.02& 28.86\\
  HJs:& 1.75& 21.91& 13.81& 1.17& 0.03& 38.67\\
\hline
Total:& 3.06& 38.26& 24.12& 2.05& 0.05& 67.53\\
B\&G:& 4.40& 28.60& 23.50& 5.80& 0.33& 62.90\\
\hline
\end{tabular}}
\end{center}
\end{table}

The number of habitable\footnote{By habitable, I mean a planet in a circular orbit whose semimajor axis is scaled so that it will have the same blackbody equilibrium temperature as the Earth.} Earth-like planets that Kepler will detect is another quantity of interest. Using the KIC, and assuming no red-noise and that every star posses a habitable Earth, Kepler should detect 36.17 planets. This is slightly less than half the number predicted in B\&G, which also assumes that every star possess an Earth, that there is no red noise in the data, and predicts 78.89 detections. The difference is attributable to the estimated spectral distribution of the KIC stars, which is skewed more towards earlier-type stars than B\&G. The skew results from the low $log(g)$ cut-off, which would select some slightly evolved stars. 

In any event, it should be stressed that both predictions (KIC and B\&G) are extreme upper limits to the number of habitable Earths that Kepler will detect. In reality, it is improbable that every star will have a habitable Earth, and that Kepler will have no red-noise in its data. Indeed, B\&G considers the effect of varying amounts of red-noise on Kepler's habitable planet detections, and finds that even small amounts could have potentially large effects on the final detection numbers. 

\section{Conclusion}

The field of transiting planets is transitioning from pure discovery, wherein every new transiting planet is newsworthy, to the details. Individual planets, aside from the smallest ones being discovered, are becoming less interesting in and of themselves, and more interesting for what they can tell us about the statistical properties of planets. Given that transiting planets are so far one of the only ways that we may infer the radius of extrasolar planets and their exact orbital properties, the statistical characteristics of this group of objects is one of the only foreseeable ways that the areas of planetary interiors, system dynamics, migration, and formation will acquire more data. 
 
That being said, it is often hard to draw specific statistical conclusions from the results of the transit surveys. In B\&G we were able to adopt hard cut-offs on our simulated detections, but the transit surveys typically do not adopt the strict detection criteria that we have used. Promising planet candidates are often followed up even if they are beyond the stated S/N or magnitude limits of the survey. Understanding and quantifying how the survey teams select candidates is vital to appropriately deriving the statistical properties of extrasolar planets. Indeed, the results of calculations shown here and in B\&G demonstrate that the actual predictions depend crucially on the specific magnitude and S/N threshold used.

In the future, besides using the results of B\&G to design more efficient transit surveys, I would hope that a more systematic approach towards transit surveys will allow the model to be used to make more specific statistical comparisons. As the number of known transiting planets grows, we would also expect that the model will be used to test different distributions of planet frequencies, periods, and radii against those observational results. This will allow us to better understand the statistics of extrasolar planetary systems, improve our ability to find new planets, and help to understand the implications of the ones that have already been detected.

\end{document}